ESSAY

# *PERMANENCE* – AN ADAPTATIONIST SOLUTION TO FERMI'S PARADOX?

Milan M. Ćirković

*Astronomical Observatory of Belgrade*

*Volgina 7, 11160 Belgrade, Serbia and Montenegro*

*e-mail:* mcirkovic@aob.aob.bg.ac.yu

**Abstract.** A new solution of Fermi's paradox sketched by SF writer Karl Schroeder in his 2002. novel *Permanence* is investigated. It is argued that this solution is tightly connected with adaptationism – a widely discussed working hypothesis in evolutionary biology. Schroeder's hypothesis has important ramifications for astrobiology, SETI projects, and future studies. Its weaknesses should be explored without succumbing to the emotional reactions often accompanying adaptationist explanations.

> The work of the Spirit of earth, as he weaves and draws his threads on the Loom of Time, is the temporal history of man as this manifests itself in the geneses and growths and breakdowns and disintegrations of human societies; and in all this welter of life and tempest of action we can hear the beat of an elemental rhythm whose variations we have learnt to know as challenge-and-response, withdrawal-and-return, rout-and-rally, apparentation-and-affiliation, schism-and-palingenesia. This elemental rhythm is the alternating beat of Yin and Yang; and in listening to it we have recognized that, though strophe may be answered by antistrophe, victory by defeat, creation by destruction, birth by death, the movement that this rhythm beats out is neither the fluctuation of an indecisive battle nor the cycle of a treadmill.

> Arnold J. Toynbee, *A Study of History*, Vol. I, Chapter III, (4) [1]

We use our intelligence to investigate the issue of hypothetical intelligences elsewhere in the Galaxy—so much is uncontroversial. Is it conceivable, however, that exactly this obvious and unavoidable selection effect causes systematic errors in our judgment on the perennial problem summarized in the famous Fermi's question: *Where is Everybody?* This is not to indicate—as many SETI-detractors have indeed done—that the search for extraterrestrial intelligence is misconceived or founded on false premises; instead, we ask a deeper question about the **intelligibility of our very concept of intelligence**. Without it, we are left in the strange position of the ancient Chinese philosopher who concluded that, since nobody knows what a unicorn really is, he might have already seen a unicorn without noticing it. In this essay, we shall briefly investigate such a proposal in the modern astrobiological context.

Fermi's question has recently become more pertinent than ever. For the first time in the millennia-long history of speculation on extraterrestrial life, in the last couple of years we got the numerical hold on the age distribution of possible life-bearing sites in the Galaxy. Seminal results of Lineweaver and his collaborators [2,3] show that Earth-like planets began forming in the Milky Way about 9.3 Gyr ago, while their average age is 6.4 ± 0.9 Gyr. This is significantly larger than the age of Earth (measured to be 4.56 ± 0.01 Gyr [4]), indicating that the difference between evolutionary ages of other biospheres in the Galaxy and ours should—on the Copernican assumption of our average location and properties—be more than a billion years. It becomes then especially hard to answer the question why we do not perceive any manifestations of Galactic supercivilizations, more than a billion years older and unimaginably more advanced than we are. A billion years ago, very simple organisms, like bacteria and acritarchs, were the only inhabitants of our planet; shouldn't we be like them to an **average** extraterrestrial intelligent community in the Milky Way? What about those which are even more advanced than the average? What about **the first** Galactic civilization? It is becoming increasingly difficult to assert that conditions on Earth, terrestrial life and intelligence are typical, or average, in the Galactic context.

One would think that at a time when books with titles such as *Fifty Solutions to Fermi's paradox* [5,6] are written and widely read, it seems nearly impossible to invent a new solution to the old puzzle of the apparent absence of extraterrestrial intelligences in our past light cone. However this is exactly something which Karl Schroeder does in his brilliant SF novel *Permanence* [7]. This is a **novel** solution in both senses of the word—it blends discoursive philosophical and scientific thinking, with a poetic expression appropriate for its format, and it has not been seriously analysed in the astrobiological research literature so far. It is not entirely surprising that a serious scientific hypothesis is formulated, in a qualitative manner, in the recreational context of a piece of SF art; astrobiology is perhaps uniquely positioned to exert such influence upon human minds of various bents. After all, much of the scientific interest in questions of life beyond Earth in the XX century was generated by works such as Herbert G. Wells' *War of the Worlds*, Sir Arthur Clarke's *2001: Space Odyssey*, or Sir Fred Hoyle's *The Black Cloud*.

And this is not the only reason to devote the present essay to analysis of this solution. As we shall see, this solution may point in the direction of a multidisciplinary synthesis—one which has both the virtue of being connected with what we know from terrestrial evolutionary biology and which could yield testable scientific hypotheses. The aim of this paper—it cannot be overemphasized—is not to selectively promote Schroeder's scenario as **the** solution of Fermi's paradox; in our present state of ignorance that would be a disservice to serious astrobiological endeavor. (This can be said without reference to inherent flaws which, in the opinion of the present author, are clearly visible in Schroeder's scenario.) However, the ideas exposed in *Permanence* certainly deserve a serious scrutiny as well as wider discussion in the rapidly expanding realm of astrobiology and SETI studies—and it is along these lines that the present essay finds its *raison d'être*.

\*  \*  \*

The plot of Schroeder's novel revolves around the dichotomy between so-called lit worlds (i.e. those surrounding Sun-like stars) and the worlds of the "Cycler Compact", surrounding brown dwarfs. Since the average distance between brown dwarfs is most probably by a factor of a few smaller than the average interstellar distance, it makes sense for the Cycler worlds to communicate via slow interstellar vessels ("cyclers"), which cruise along a preset path, using the Galactic magnetic field to maneuver. Usage of cyclers creates a specific culture (of "halo worlders"), which is beautifully described by Schroeder.

*Permanence* begins auspiciously enough. We join the protagonist, Rue Cassels on the space station that is her home. It is located near a brown dwarf star which belongs to the Cycler Compact. With the invention of an FTL drive that only works around bright or "lit" stars, the economy based on these ships fades away. All this is important background of the story, but for now Rue is escaping the station and her abusive brother with a ship. On the way to a planet around the dwarf, she lays claim to something. The action then flashes forward to the planet around the dwarf star where Rue meets her cousin. Soon the book jumps again to join a distinguished astrobiologist, Dr. Herat, his companion, Michael Bequith, and some military-types who study alien civilizations and their remains. They join up with Rue and her companions, and a story finally begins to form centered on the basic "big dumb object". There are a few more jumps in the story, but the latter two-thirds of the book maintain a reasonable narrative coherence.

All of the classical elements of great science fiction are here—from action and adventure to fading and fallen civilizations, from Galactic politics and intrigue to outright war and rebellion. What we are interested here, however, is the underlying astrobiological premise, which contains the best elaboration so far of an original solution to Fermi's paradox. There are few other books of fiction so thoroughly infused with the astrobiological topics and issues. Even the fortuitous or vaguely symbolic subjects, like the title of Part One—"Ediacara"—have astrobiological significance.[1]

---

[1] Ediacaran fauna—itself named after Ediacaran Hills in the outback of Southwestern Australia—presented the last (Vendian) phase of the Precambrian period. Recent radiometric studies fixed

Schroeder's astrobiologists of the future endorse part of the "rare Earth" paradigm of Ward and Brownlee [9], notably the ubiquity of simple life:

> Enough infrared leaked out of Erythrion to heat the surface of Treya to livable temperature, and tidal and induction heating kept it volcanically active. But without a sun, life had never developed here—or rather it had developed and died out a number of times... (p. 27)

However, as we shall see, Schroeder radically departs from the "rare Earth" paradigm, and his solution is, in a sense, polemic with Ward and Brownlee.

<p align="center">*　　*　　*</p>

Schroeder starts from the physical background assumed by most researchers, and the one in which Fermi's paradox is most sharply manifested: his Galaxy is continuously habitable for most of its history, or at least for periods of time much longer than the timescale for the rise of a starfaring technological species.[2] Why, then, astrobiological investigations of several centuries have not revealed advanced technological species similar to humans? It is crucial to keep in mind that this is not just a fictional question in a fictional scenario; **it is our present dilemma in SETI studies when faced with Fermi's paradox,** only scaled up a bit. Schroeder hopes that this scaling, this powerful magnification lense, will give us a glimpse of the solution. In order to achieve this, he needs to delineate two mutually exclusive philosophies underpinning our understanding of the role of intelligence in the universe.

their first appearance at 565 million years before the present, and they have lasted till the Cambrian boundary, about 543 million years ago. They are now thought to represent transitional intermediates to the Cambrian animals—a kind of fuse on the famous Cambrian Explosion. For more details see, e.g., [8].

[2] E.g., Schroeder's civilization of Dis which plays a crucial role in humans' understanding is about 3 billion years old.

Schroeder first briefly sketches what he calls **providential** view of astrobiology. It seems to be prominent today in the actual world of astrobiological and SETI researchers:

> The truth is that we are intelligent animals, but animals just the same, subject to the inescapable laws of our evolution. Our first theories about alien intelligence were *providential*: we believed with Teilhard de Chardin, that consciousness is a basic characteristic of complex thinking entities. When we developed the FTL drive, we burst into the galaxy in search for beings more "evolved" than ourselves, in the belief that a universal Reason would unite us with other species at the same level. (p. 108)

This view has been criticized by a number of influential critics in both scientific and literary domains, including Ernst Mayr, George G. Simpson, John Barrow and Frank Tipler, Stanislaw Lem, Ronald Bracewell, Sir Arthur Clarke, Strugatsky brothers, Zoran Živković, and others, though it is still predominant in the SETI field.[3]

But what should such a view be contrasted with? Schroeder offers a strong alternative, which he does not dub, but which we can, for reasons to be shortly described, call **adaptationist**:

> What we found instead was that even though a species might remain starfaring for millions of years, consciousness does not seem to be required for toolmaking. In fact, consciousness appears to be a phase. No

---

[3] This is somewhat ironical, since most of actual SETI researchers are agnostic, at least from an external perspective, and would be surprised to find their views associated with Teilhard de Chardin, who in scientific circles earned—perhaps undeservedly—a reputation of being a quasi-scientist and mystic. In general, religious views are nowadays usually associated with **opposition** to SETI and contact-scepticism (a prominent example is Frank Tipler; see [10,11]). Even more to the point, Teilhard himself did not believe in extraterrestrial intelligent life (at least until the last phase of his thinking, in 1950s, when he seemingly revised some tenets of his system); in *The Phenomenon of Man* he wrote that evolution of intelligence elsewhere has a "probability too remote to be worth dwelling on" [12]. However, on a deeper level, Schroeder's characterization is correct. As succinctly put in Wildiers' foreword to the book of Hague [13]: "As a student of the phenomenon of man, Teilhard de Chardin constantly refused to see in reflective consciousness a mere epiphenomenon, a mere accident thrown up by nature, unrelated to the underlying structure of our universe." (p. 7)

> species we have studied has retained what we could call self-awareness
> for its entire history. Certainly none has evolved into some state *above*
> consciousness. (p. 108)

This is the crux of the problem (for the astrobiologist protagonists of the novel; solution for us, willing to resolve Fermi's paradox): our estimates and expectations of the phenomenon of intelligence—which is, above all, a biological phenomenon—are wrong. Intelligence is significant only insofar as it offers an evolutionary advantage, a meaningful response to the selective pressure of the fluctuating environment. Only so far, and no further is the "selfish gene" [14] willing to carry that piece of luggage.

This approach to explanation in biology is known as **adaptationism**; its major proponents being distinguished biologists such as Richard Dawkins and John Maynard-Smith, as well as contemporary philosophers such as Daniel Dennett or Eliot Sobber. Adaptation is a trait that has been selected for by natural selection. Adaptationist hypothesis can be conventionally defined as a statement that asserts that a given trait in a population is an adaptation. In other words, natural selection is the major (if not the sole) cause of presence and persistence of traits in a given population. The definition of Sober in his influential book [15] goes as follows:

> Adaptationism: Most phenotypic traits in most populations can be
> explained by a model in which selection is described and nonselective
> processes are ignored. (p. 122)

Examples of adaptationist explanations abound. Camouflage colors of birds and insects, Eskimo faces, horns of *Ontophagus acuminatus,* and myriads of other observed properties of living beings are interpreted as giving their carriers an advantage in the endless mill of natural selection. Their genes are more likely to propagate along the thousands of generations of natural history. The most extreme version of adaptationism is sometimes called gene-centrism and is expounded by Richard Dawkins, and neatly encapsulated in the title of his best-selling book *The*

*Selfish Gene* [14]: genes are using (in a sufficiently impersonal sense of the word, see Ref. 16) organisms to propagate its own copies as efficiently as possible in time. It may be of historical interest that adaptationism is usually traced back to Alfred R. Wallace, one of the two great biological revolutionaries, who was also one of the forefathers of modern astrobiology with his intriguing and remarkably prescient 1903 book *Man's Place in the Universe* [17].[4]

This view is the scientific foundation of Schroeder's solution to Fermi's paradox. Intelligence is an adaptive trait, like any other. Adaptive traits are bound to disappear once the environment changes sufficiently for any selective advantage which existed previously to disappear. In the long run, the intelligence is bound to disappear, as its selective advantage is temporally limited by ever-changing physical and ecological conditions.

> Look at crocodiles. Humans might move into their environment—underwater in swamps. We might devise all kinds of sophisticated devices to help us live there, or artificially keep the swamp drained. But do you really think that, over thousands or millions of years, there won't be political uprisings? System failures? Religious wars? Mad bombers? The instant something perturbs the social systems that's needed to support the technology, the crocodiles will take over again, because all they have to do to survive is swim and eat.

Schroeder's protagonist continues:

> It's the same with consciousness. We know now that it evolves to enable a species to deal with unforeseen situations. By definition, anything we've mastered becomes instinctive. Walking is not something we have to consciously think about, right? Well, what about physics, chemistry, social engineering? If we have to think about them, we haven't mastered them—they are still troublesome to us. A species that succeeds in really mastering something like physics has no more need to be conscious of it. Quantum mechanics becomes an instinct, the way ballistics already is for us. Originally, we must have had to put a lot of thought into throwing things like rocks or spears. We eventually evolved to be able to throw without thinking—and that is a sign of things to come. Some day, we'll become... able to maintain a technological infrastructure without needing to *think* about it. Without need to think, at all...

---

[4] Ironically enough, Wallace also (in)famously insisted—to Darwin's dismay—that human mind shows so many nonadaptive features, like the music appreciation, abstract mathematics or spiritual communication, that it is a sure sign of an intervention by higher intelligence. See, e.g. [18].

The idea that intelligence might one day cease to be **useful** to its possessors has been present in SF and popular culture for a long time, at least as a sort of black humor joke. We shall consider some of the similar ideas below. However, nobody has gone in that direction so far and so consequently as Schroeder. He envisions an incredibly advanced (by human standards) culture facing this frightening evolutionary impasse:

The inhabitants of Dis studied previous starfaring species. The records are hard to decipher, but I found evidence that all previous galactic civilizations had succumbed to the same internal contradictions. The Dis-builders tried to avoid their fate, but over the ages they were replaced on all their worlds by fitter offspring. These descendents had no need for tools, for culture, for historical records. They and their environment were one. The conscious, spacefaring species could always come back and take over easily from them. But given enough time... and time always passes... the same end result would occur. They would be replaced again. And so they saw that their very strength, the highest attainments they as a species had achieved, contained the seeds of their downfall.

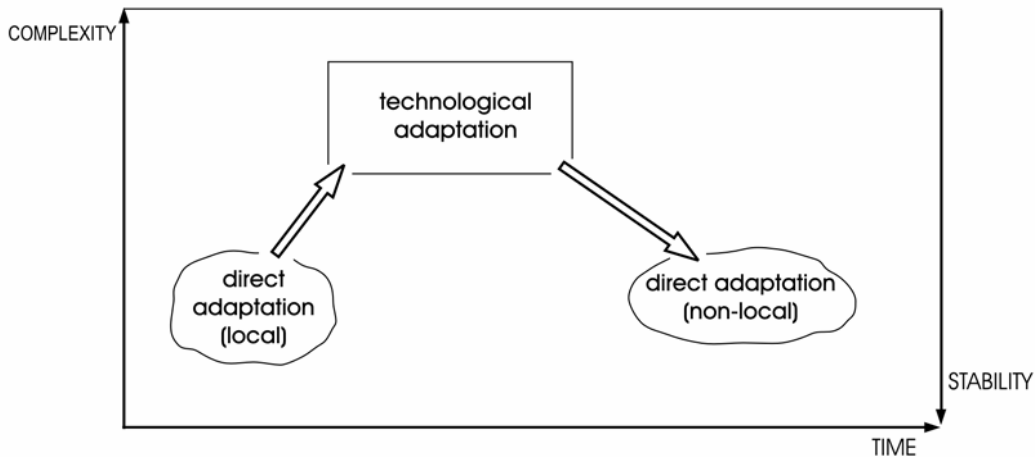

**Figure 1.** A schematic rise and fall of technological civilizations in the Galaxy according to Schroeder's picture. Key concept is adaptation.

This chain of events is schematically shown in Figure 1. An intelligent species can last long in the state of direct adaptation to their environment on the home planet (local)—several hundred thousand years in the case of *homo sapiens sapiens*. A rather fast transition from direct to technological adaptation corresponds to building of a technical civilization, this crucial ingredient of all SETI studies. But the stage of technological adaptation, distributed all over the various planets, is inherently less stable. In a long run, it will tend to pass into a state of fragmented habitats reverting to direct biological adaptation ("crocodiles returning"). The second transition is much slower and might be occasionally interrupted or arrested; yet, the general trend toward return to direct adaptation is inescapable. This bifurcation is, in this view, tantamount to extinction of the original intelligent species, and its remains are gradually submerged into the general astrobiological "noise" of the Galaxy. *Sic transit gloria mundi*. The transient nature of the phase of technological adaptation constitutes the bulk of the "Great Filter" explaining the *silentium universi* [19,20].

There is another perspective in which we can approach the same issue. What can we, from our standpoint of predominant ignorance, conclude about other Galactic intelligent species? It'd be all too easy for us to unthinkingly assume that all intelligent life is carbon-based, metabolizes using oxidation, and eats pizza, just because folks around here do. But how to escape this unthinking parochialism without falling into some sort of skepticism? Schroeder proceeds in a way which will look familiar to both physicists and biologists: by assuming that all these observed phenomenological aspects of intelligent life are in the final analysis consequences of a deeper underlying principle, which is adaptationism. The diversity of Earth's biosphere is astonishing; and yet, even the most ardent ecologist and conservationist would hardly argue that it is but an infinitesimal fraction of possible diversity, of the "Library of Mendel"[5], which would contain all possible genotypes—even if we retain the same biochemical basis of life. (An even greater variety certainly becomes available if we allow for different biochemistries at various locales all over the Galaxy.) Thus, our eating pizza would be—in the more extreme adaptationist variants, particularly in evolutionary psychology—an external manifestation of the underlying molecular striving for accommodation to our particular environment. An incredibly advanced alien astrobiologist could, in principle, infer the existence of pizza-eating from sufficiently detailed understanding of the terrestrial physical and chemical environment. In the same manner, the wide spectrum of diverse Galactic habitats would produce different spectrum of adaptive behaviors. But the overarching logic of the adaptive development necessarily contains a boundary: one can be perfectly adapted to one's environment—but certainly not more adapted than that. And, Schroeder suggests, the perfect adaptation in the numerically by far predominant locales does not include tool-making abilities or star-faring capacities.[6]

Schroeder suggests that the ultimate Copernican revolution should bring forth the view that intelligence is not only common, on the average, but that it is in fact **unimportant**. It makes as much impact on the overall unfolding of cosmic events in the long run as does a blue color on wings of a particular species of terrestrial butterflies. To ask for anything grander, deeper, more spiritually profound, or elevated, is just a manifestation of our incurable "psychocentrism".

---

[5] See Dennett, Ref. 16.

[6] Astrobiology seems to be particularly suitable grounds for exercising adaptationism, since it by definition includes most varied environmental conditions and selective pressures possible. On the other hand, this will make the explanatory task much more difficult and trap-ladden. As Griffiths [21] wrote: "The adaptationist assumes that almost all forms are developmentally possible, so learning that the actual form is possible does not explain the contrast between this form and the adjacent forms."

And, in the final analysis, psychocentrism just steps in shoes of "classical" anthropocentrism, either in its old-fashioned teleological or modern (e.g., "rare Earth") form.

This is nicely encapsulated in an ancient cartoon by Henry de la Beche (under the wonderful title "AWFUL CHANGES") shown in Figure 2, and widely popularized by Stephen Jay Gould in his *Time's Arrow, Time's Cycle* [22]. Resurrected ichthyosauri of the far future of the Earth muse over human fossil remnants, concluding that this creature could not be well adapted to its environment. Abstract for the moment the obvious fact that ichthyosauri would have needed intelligence and even tool-making in order to give lectures on paleontology; this is just the superficial problem of the situation. The deeper issue—and one which cuts to the core of the adaptationist approach to Fermi's paradox—is that those terrifying sea reptiles from Mesozoic **need not** know anything about intelligence or pizza-eating or spaceship construction whatsoever; taken to the adaptationist extreme (often ridiculed by Gould himself), they could even conclude that the fact that humans went extinct is the very proof of their inability to cope—of their poor adaptive qualities, that is. This reasoning does possess an element of circularity, but as Raup laments [23,24], this seems to be the case with most of **our** accounts of extinction of earlier species! With exception of the dinosaurs and the small subset of other species which perished, we rather confidently state, in sudden catastrophes having nothing do to with their adaptation levels, our view of most of the extinctions (especially those "background" ones, constituting a reference point for defining "mass extinction episodes") is characterized by the same circularity: in accordance with the Darwinian view, we are forced to conclude that they were poorly adapted to their environment and **thus** became extinct. Nothing is said about intelligence or capacity for pizza-eating, as well as of any other **particular** trait.

Why none of the species have evolved to a state above consciousness, something like much speculated "transcedence" figuring (usually clumsily) in so many other SF works? The answer in the context of the adaptationist paradigm is clear: because there is no environmental feature which would require anything above the consciousness. Even the consciousness itself is useful only in the short run!

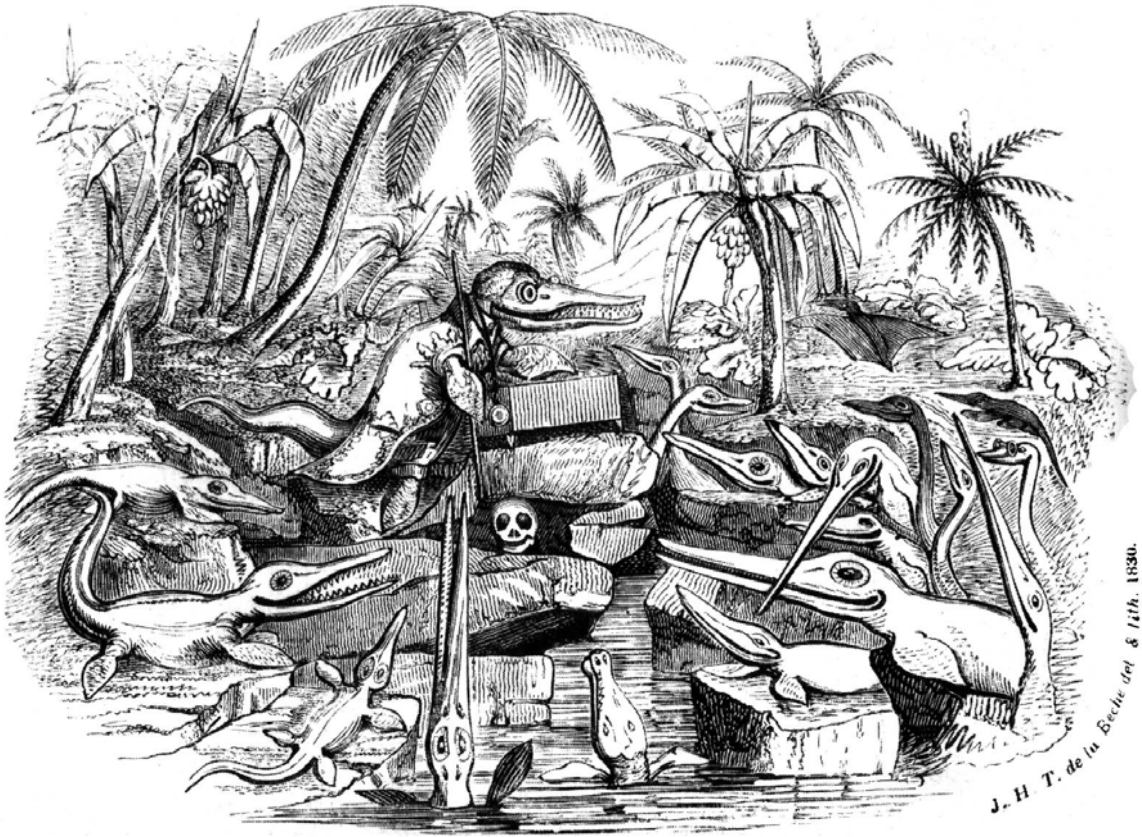

**AWFUL CHANGES.**

MAN FOUND ONLY IN A FOSSIL STATE.——REAPPEARANCE OF ICHTHYOSAURI.

*A Lecture.*—" You will at once perceive," continued PROFESSOR ICHTHYOSAURUS, " that the skull before us belonged to some of the lower order of animals; the teeth are very insignificant, the power of the jaws trifling, and altogether it seems wonderful how the creature could have procured food.'

**Figure 2.** A wonderful 1830 cartoon by Henry de la Beche, resuscitated by Stephen Jay Gould in *Time's Arrow, Time's Cycle* [22]; see the historical discussion of Rudwick [30]. A powerful predator like ichthyosaurus—revived in far future, according to a bizarre speculation of Charles Lyell which is satirized here—finds hard to conceive that long-ago extinct humans had ever been well-adapted to their environment.

Simplified variations or hints on the same theme have appeared from time to time in both scientific and popular literature. Interestingly enough, both Schroeder (private communication) and Dennett [16] find a strong inspiration for their adaptationist views in Friedrich Nietzsche's philosophy. The very concept of "permanence" has important similarities with Nietzschean "eternal return of the same". While this topic is certainly too complex to be investigated within the scope of the present essay, we shall briefly return later to another similarity between Schroeder's and Nietzsche's worlds – their overarching and unavoidable nihilism.

The picture of bifurcation of species due to adaptation (although intentional in this case) is the major motiff of the classic *The Seedling Stars* by James Blish [25]. However, the darker aspect of the situation in which the limits of knowledge and intelligence are reached is only rarely shown in the SF discourse. Such fate befells the flabbergastingly old and advanced civilization of Transmuters in Greg Egan's *Diaspora*:

> 'Then why did they stop?'
>
> 'Because there was nothing more to do. ... They'd seen everything they wanted to see in the outside world – they'd risen through at least six universes – and then they'd spent two hundred trillion clock ticks thinking about it. Building abstract scapes, making art, reviewing their history... We'll never decipher it; we'll never know for sure what went on. But we don't need to. Do you want to ransack the data, hunting for secrets? Do you want to rob their graves?' (Ref. 26, p. 358)

A suggestion along these lines has been made by the Russian author Vladimir Khlumov in 1987 [27] in a short story; it is somewhat elaborated in a discoursive form by the distinguished astrophysicist Vladimir Lipunov [28]. Khlumov and Lipunov suggest that there is nothing inherent in the concept of intelligence which can be manifested without reference to the empirical content of the world. Thus, if the amount of information describing this empirical content in its entirety is finite, the intelligence is finite as well. What happens when this amount of information is processed? Intelligence ends as well, reply Klimov and Lipunov. When every problem is solved, when no unanswered question arises, when no new theory is ever necessary for explanation of any phenomenon, it seems natural enough to suppose that the mill of intelligence grinds to a halt. Is the low-complexity of the Universe (and the suggestion often enjoyed in some circles of theoretical physicists that the "Theory of Everything" is at hand) the reason why we do not perceive advanced extraterrestrials?

Perhaps the most pertinent scientific account pointing in this direction is a brief paper given by David M. Raup, one of the leading paleontologists of XX century, at 38th International Astronautical Congress at Brighton, UK in 1987, under the indicative title *Unconscious intelligence in the universe* [29]. Raup has been, for much of his illustrious career, one of the major contributors to our understanding of the great episodes of mass extinctions of species in Earth's past, as well as a champion of the extraterrestrial causes for at least some of those (e.g. [23,24]), thus exemplifying a multidisciplinary character of the astrobiological enterprise. In this particular contribution, he speculates that animals on other planets may have evolved, by natural selection, the ability to communicate by radio waves. Our own SETI projects are still mostly radio-based. Radio communication in such non-intelligent organisms as proposed by Raup would persist much longer than radio communication developed by intelligent beings, which would be ephemeral due to cultural changes. Strategies for SETI should take the possibility of such radio communication into account.

The link between Raup's and Schroeder's scenario becomes clear when we consider the adaptationist grounding of both. If we recognize that something we parochially regard as the summit of intentional intelligent tool-making—radio communication, that is—can evolve by direct adaptation, which other feature of technological civilization is safe from emulation by Nature? Raup emphasizes the long-term aspect of the situation, and that the adaptive radio sources can be predominant in the total set of radio sources. Similarly, in Schroeder's picture, the number of Galactic habitats with direct adaptation (either before or after the technological phase) is much greater than the current number of sites of technological civilization.

It is far beyond the scope of this essay to even remotely discuss pros and cons of adaptationist doctrine as such. It is enough to be aware of the cloud of polemic and debate it is surrounded with for almost a century and a half since its inception by Darwin (or, more to the point, Wallace). It is to be assumed that the debate will continue in the future, and quite possibly will remain with us when first sites of extraterrestrial life are found. Keep in mind that this adaptationist solution is not—at least on the human timescales!—what Dennett rightly condemns as "greedy" adaptationism. It does not assume (with, for instance, Skinner's behaviorism) that culture and its all achievements, fine art, letters, etc. are just highly specialized adaptations and/or can be **determined** by biological mechanisms.[7] This strong thesis (or caricature) is unnecessary to begin with; the solution to Fermi's paradox simply requires that **in the long run** culture is **irrelevant** to the history of species as a whole. The issue of relevancy is entirely separable from the issue of biological (or other) determinism. This division perhaps does not remove our emotional revulsion at such reductionist approach, but blunts criticisms by anti-adaptationists. In fact, Schroeder offers many examples that the culture is in fact **not** determined by "genetic fate" in store for a species; for instance, Dr. Herat's report which explains this fate continues:

> The Panspermia Institute was formed out of the disappointment of this discovery. We sought to uncover the conditions that give rise to sentience; if we could not find aliens like ourselves, perhaps we could guide candidate species into our mode of experience. (p. 108)

This external (in Plato's sense) teleological engineering of candidate species would be antithetical to the default assumption of adaptationism. Herein, parenthetically, lies one of the weaknesses of Schroeder's solution, since there is no proof—or indeed a clear counterargument—that a chain of contingent intentional strategies for preventing adaptationist devolution could not extend over an arbitrarily long time in a history of advanced civilization. In other words, things certainly can and will go wrong at many locations over a sufficient period of time, but one could easily imagine planning and building of "concentric rings" of safety mechanisms, each activated after all previous have failed. Each mechanism individually can and will eventually fail, but there seems to be no clear reason why the entire system could not be simply vast enough and continuously assembled over time in order to counter the diverging trends of isolation and devolution.

---

[7] Another instance of such greediness is the (in)famous example of Aztec ritual cannibalism rightly mocked by Gould and Lewontin in the *Spandrels* paper [31].

Another important feature of Schroeder's solution is that, contrary to what astrobiologists and SETI researchers usually assume, the **speed** of evolution in a given locale is unimportant for determination of the number and age of accessible alien civilizations. It has been usually assumed that fast evolution (similar to the techno-evolution we perceive on Earth in last several centuries) will cause civilizations to evolve up the ladder of Kardashev types: from Type I (more or less similar to present-day human) to Type II (capable of building a Dyson sphere and marshalling all resources of its planetary system) to Type III (pan-galactic civilization spanning many planetary systems and managing resources on the galactic scale). Schroeder dispells this illusion: in his picture, faster the evolution, faster is the devolution of a species into the non-conscious state. Standing next to an alien artifact, a protagonist muses (p. 217):

> Maybe this really was just a thing, magnesium alloy and aerogel filling, no more or less significant than any rock. It had been created by blind evolutionary fate, as had he; he wasn't going anywhere but where his genes led him; nor was the human race going anywhere. Herat had proven that—they were at the top of the evolutionary arc, with nowhere to go but down. So all this investigation was futile. You could already see the seeds of decay...

(Schroeder parenthetically answers another pertinent question which has probably puzzled more than one SF writer—is there a credible threat to survival of an advanced interstellar society spanning many planetary systems? Contrary to the juvenile naive picture of grand cosmic catastrophes, the true threat lurks deep within the very essence of intelligent beings. As the Bard wrote: *The fault, dear Brutus, is not in our stars // But in ourselves, that we are underlings.*)

What adaptationism fails to explain on Earth, it will fail to explain in the Milky Way. Perhaps the foremost problem with the adaptationist doctrine as currently presented by the evolutionary orthodoxy is its failure to be useful in cases of brief and sudden episodes in the history of life known as the mass extinctions. In the words of Raup's title [23], in such times "luck" is more important than "genes". In our (highly incomplete) fossil record, "Big Five" mass extinction episodes stand as the most remarkable features, after the Cambrian Explosion, of the history of life. We cannot, of course, delve into this huge topic here in detail, but it is important to notice that the general assumption that mass extinctions have decidedly impacted the vector of evolutionary progress on Earth is probable to hold in the wider astrobiological context of the Milky Way either. Of course, the global regulation mechanism in the galactic case must be very different, but a hopeful candidate exists in shape of gamma-ray bursts (Ref. 32 and references therein). The true solution to Fermi's paradox will be, one is tempted to suspect, the one which could strike a fine balance between the catastrophist and gradualist effects in the course of time—a sort of "punctuated equilibrium" [33] on the Galactic scale!

<p style="text-align:center">*    *    *</p>

Finally, a word about our emotional reactions to this sort of solution might be in order. Schroeder's is a depressing hypothesis, in a sense even worse than the conventional "catastrophic" scenarios in which intelligence self-destructs with a boom. It is worse than self-destruction, since the latter at least pinpoints a **peculiar feature** of intelligence and its technological post-evolution; it subconsciously affirms the awesome **power** of intelligence, even if misused. But Schroeder gives no quarter here: there can be nothing peculiar about intelligence, at least nothing more peculiar or exceptional than the color of a butterfly's wings. Its physical power of creation or destruction is completely ephemeral and—ultimately—irrelevant.

It is difficult to avoid being revolted by Schroeder's suggestion. But this is also old news, since it is in the same manner that adaptationist explanations in other fields of biology and life sciences are often emotionally unpalatable and even revolting. Disciplines such as sociobiology and evolutionary psychology have sprung out of application of the adaptationist reasoning to animal and human behavior. Their very existence sparks controversy to this day, but the controversy can be construed as a resistance of the same "psychocentrism" we grow accustomed to. Is the rejection not only of any value we may find in our lifeworks, but also the totality of the "civilization", "culture", and other similar memes we use to denote long-term values we hope to create on the literally cosmic scale a caricature or even a *reductio* of the adaptationist doctrine or just its logical and sober consequence? The future of our own species will answer this question, in action if not in discourse, but it **will** be answered at some point.

All of this suggests a reinterpretation of Karl Schroeder's novel title as a monumental irony: while evolutionary change is ubiquitous in the universe, intelligence itself is anything but permanent. Its wakes and tides are marked feature of the Galaxy in the astrobiological context, but we shall have to adapt (no pun intended!) to the Heraclitean fact that only change is permanent.

**Acknowledgements.** Kindest thanks go to Anders Sandberg who was the first to point out the relevance of Schroeder's work to me one sunny July 2003 morning in Novi Sad, and to Dušan Indjić—Luigi for friendly help with figures. Other technical help has been received from Vjera Miović, Maja Bulatović, Alan Robertson, Nikola Milutinović, Vesna Milošević-Zdjelar, Samir Salim, Branislav K. Nikolić, Srdjan Samurović, and Nick Bostrom. This is an opportunity to thank *KoBSON* Consortium of Serbian libraries, which enabled at least overcoming of the gap in obtaining the scientific literature during the tragic 1990s. Pleasant discussions of the relevant topics with Ivana Dragićević, Nick Bostrom, Mark A. Walker, Robert J. Bradbury, Zoran Živković, Petar Grujić, G. David Brin, Zoran Knežević, Slobodan Popović, and Richard Cathcart have helped improve this manuscript. The author has been partially supported by the Ministry of Science and Environment of Serbia through the Project #1468, "Structure and Kinematics of the Milky Way." Any eventual merits in this work are due to Irena Diklić, whose curiosity, tenderness, and kind support have presented a constant source of inspiration and encouragement during the work on this project.